 \documentclass{aastex}
 \usepackage{emulateapj5}

\newcommand{\gsim}{\;\lower4pt\hbox{${\buildrel\displaystyle >\over\sim}$}\;}
\newcommand{\lsim}{\;\lower4pt\hbox{${\buildrel\displaystyle <\over\sim}$}\;}

\font\eightrm=cmr8 scaled 1000
\font\sevenrm=cmr7 scaled 1000

\slugcomment{To appear in the Astrophysical Journal Letters }

\shorttitle{Circumnuclear Spirals and Starburst Rings}
\shortauthors{Lou, Yuan, Fan, Leon}

\begin{document}


\title{Circumnuclear Spiral Arms and Starburst Rings\\
in Magnetized Barred Spiral Galaxies}


\author{Yu-Qing Lou \altaffilmark{1,\hbox{ }2}, \ \ 
Chi Yuan \altaffilmark{2}}

\author{Zuhui Fan \altaffilmark{1,\hbox{ }2}, \ \ 
Stephane Leon \altaffilmark{2} }

\affil{$^1$Department of Astronomy and Astrophysics, The Univ. of Chicago,
    5640 S Ellis Ave, Chicago, IL 60637 }
\email{lou@oddjob.uchicago.edu}
\affil{$^2$Institute of Astronomy and Astrophysics, Academia Sinica,
P.O. Box 1-87, Nankang, Taipei, Taiwan 115}



\begin{abstract}
The Seyfert galaxy NGC 1097 has 
an extended neutral hydrogen disk, a companion,
a prominent bar and a luminous circumnuclear starburst ``ring". 
Magnetic fields as revealed by nonthermal radio-continuum 
emissions correlate well with the optical barred spiral structure 
on large scales, have a gross enhancement overlapping with the 
optical/infrared ``ring", and show a trailing swirl around and 
within the ``ring". We propose a scenario of bar-excited 
long-trailing fast magnetohydrodynamic (MHD) density waves at the 
modified inner Lindblad resonance (mILR), physically identified 
with the outer rim of the ``ring". These sustained outgoing 
long-waves are bounced back by the $Q_M$-barrier in the form of 
incoming short-trailing waves. The damping of these waves deposits 
a {\it negative} angular momentum into the magnetized circumnuclear 
gas disk. Thus, gas materials spiral inward, bring in frozen-in 
magnetic flux, and accumulate inside the mILR to create a circular 
zone of high density and magnetic flux vulnerable to massive star 
formation. Depending on the wave damping efficiency, this process 
may simultaneously sustain a net mass inflow across the ``ring" 
and toward the nucleus. A wavelet analysis on a {\it Hubble Space 
Telescope} image of central NGC 1097 shows a distinct two-arm spiral 
structure extended down to the nucleus as a strong evidence for 
circumnuclear MHD density waves. 
We predict that magnetic-field observations with improved 
sensitivity and resolution would reveal a specific correspondence 
between circumnuclear optical and magnetic field spirals much as 
those known to exist on large scales in nearby spiral galaxies, 
including NGC 1097. 
\end{abstract}




\section{Introduction}

In barred spiral galaxies, luminous circumnuclear rings of star 
bursts (Buta 1986; Telesco 1988; Barth et al. 1995) commonly 
appear. Some galaxies with active nuclei (AGNs) are also known 
to contain circumnuclear starburst rings (Simkin et al. 1980; 
Arsenault 1989; Barth et al. 1995). The origin of such 
circumnuclear activities, monitored in a wide spectrum 
including optical, infrared and radio-continuum bands etc, 
remains to be understood.
(Lynden-Bell 1969; Shlosman et al. 1990). 
Even in the absence of self-gravity and magnetic
field, numerical simulations (Piner et al. 1995) have produced 
gross features of circumnuclear rings in a fixed bar-potential. 
While the driven flux of angular momentum is independent of
the details of the disk modeling, the exclusion of density waves 
that involve self-gravity will however affect the removal efficiency 
of disk angular momentum and the 
absence of magnetic field would at least make the formation of 
{\it massive} stars in the circumnuclear ``ring" unlikely 
(Shu et al. 1987; Elmegreen 1994). In this Letter, we advance a 
physical scenario for bar- or satellite-excited {\it fast 
magnetohydrodynamic (MHD) density waves} (FMDW) in a magnetized 
self-gravitating disk (Fan \& Lou 1996; Lou \& Fan 1998; Lou et al. 
2001) at the modified inner Lindblad resonance (mILR) to 
illustrate the processes of spiral FMDW propagation, angular 
momentum transfer, forming circumnuclear rings, and wave-induced 
net mass accretions across the ring and toward the nucleus, and 
to comprehend optical, CO, H{\sevenrm I}, infrared and 
radio-continuum observations of the barred spiral galaxy NGC 1097 
in dynamic, morphologic, and diagnostic contexts (Rickard 1975; 
Meaburn et al. 1981; Ondrechen \& van der Hulst 1983; Hummel et al. 
1987; Gerin et al. 1988; Ondrechen et al. 1989; Barth et al. 1995; 
Beck et al. 1999; Kotilainen et al. 2000; Emsellem et al. 2001). 
While we focus on NGC 1097, the scenario is generally applicable to 
circumnuclear spirals and starburst rings in magnetized barred spiral 
galaxies (e.g., NGC 6951, NGC 2997 etc.). This conceptual framework 
also bears broad implications to accretions and MHD processes in other 
astrophysical disk systems such as a protostellar disk system. 

With a {\it high} stellar velocity dispersion 
($\gsim 100\hbox{ km s}^{-1}$) and a {\it low} gas temperature, 
the central circumnuclear region of a spiral galaxy contains a 
thin stable\footnote{Stability against bar-type instabilities 
is meant. Starbursts involve gravitational collapses on 
subscales along an accumulated circular zone.
} gas disk 
within a few kiloparsecs (kpc). The gas rotation curve in an 
extended radial range is determined by the {\it total} mass 
distribution in the galaxy (Meaburn et al. 1981; Blackman 
1981; Gerin et al. 1988; Ondrechen et al. 1989) and the 
{\it background} magnetic field in the circumnuclear disk 
plane of a strength $\gsim$ several tens of $\mu$G (Beck et al. 
1999) is taken to be axisymmetric and azimuthal\footnote{It is an 
idealization to posit a purely azimuthal background magnetic field, 
based on which we perform analysis for spiral MHD density waves.
In the circumnuclear domain
under consideration, observed magnetic field deviations
are significant. The results of a perturbation analysis
is only indicative. The effectiveness of this approach may
be questionable in the fully nonlinear realm.}
to avoid the magnetic-field winding dilemma (Roberts \& Yuan 1970).
This circumnuclear magnetized gas disk is embedded with 
molecular clouds (Gerin et al. 1988) with a mean-free path 
$l_c$ and a velocity dispersion $v_c$. When disturbed, spiral 
FMDWs involve collectively the magnetized gas disk embedded 
with clouds. The large-scale {\it barred spiral structure 
outside} the central gas disk rotates at a pattern 
speed $\Omega_p$ and gives rise to a time-dependent 
{\it external} periodic gravitational potential $\phi^E$ 
as felt by the central magnetized gas disk.
In cylindrical coordinates $(r,\theta,z)$, our theory is 
developed using the standard two-dimensional equations for 
coplanar MHD density waves in the central magnetized gas 
sheet (Lou \& Fan 1998) at $z=0$ with $\phi^E$ varying slowly 
in $r$ as an {\it inhomogeneous driving term} (Goldreich \& 
Tremaine 1978a,b; Yuan \& Cheng 1991).

\section{Key Results of Theoretical Analyses}

Without such $\phi^E$ driving and away from the resonances, 
the dispersion relation for {\it free} FMDWs in the WKBJ 
or tight-winding regime (Shu 1970; Fan \& Lou 1996) is
$$
(\omega-m\Omega)^2\approx\kappa^2+k^2
(C_A^2+C_S^2+v_c^2-2\pi G\mu_{0}/|k|)\ ,\eqno(1)
$$
where $v_c^2$ mimics an ``effective pressure" due to 
random cloud velocities, $\omega$ is the angular frequency 
in an inertial frame of reference, $m\geq 0$ an integer for 
the number of spiral arms, $\Omega$ the disk angular speed, 
$\kappa\equiv [(2\Omega/r)d(r^2\Omega)/dr]^{1/2}$ the 
epicyclic frequency, $G$ the gravitational constant,
$\mu_{0}$ the background surface mass density, $k$ the radial 
wavenumber, $C_S$ the sound speed, and $C_A$ the Alfv\'en 
speed (Lou \& Fan 1998). The corresponding 
FMDW amplitude equation (Fan \& Lou 1999) is
$$
{d\over dr} \bigg [ {r\mu_0 k\over (\omega-m\Omega)^2}
\bigg(C_S^2+v_c^2+C_A^2-{\pi G \mu_0 \over |k|}\bigg)
|\tilde v_r|^2\bigg]=0\ ,\eqno (2)
$$
where $\tilde v_r(r)$ is the magnitude of the radial velocity 
$v_r$. Equation (1) is quadratic in $|k|$ and contains the 
familiar short- and long-branches of FMDWs (Lou \& Fan 1998). 
For FMDWs, it follows that $B_{\theta}\mu\cong\mu_{0}b_{\theta}$, 
where $\mu$ is the surface mass density perturbation, and 
$b_{\theta}$ and $B_{\theta}$ are the perturbation and background 
azimuthal magnetic fields; the enhancements of $\mu$ and 
$b_{\theta}$ are therefore in phase.

The stability against {\it axisymmetric} ring fragmentation 
requires the MHD version of Toomre's $Q$ parameter
$Q_M\equiv \kappa (C_A^2+C_S^2+v_c^2)^{1/2}/(\pi G\mu_{0})>1$
in a magnetized self-gravitating gas disk (Lou \& Fan 1998). 
The radial group velocity of FMDWs is
$$
F_G=-{\partial\omega\over\partial k}
\cong -{\hbox{sgn}(k)[(C_A^2+C_S^2+v_c^2)|k|-\pi G\mu_{0}]
\over (\omega-m\Omega)}\ ,\eqno(3)
$$
with $F_G>0$ and $F_G<0$ for outward and inward propagations. 
By our sign convention, $k>0$ and $k<0$ represent leading and 
trailing spirals. Inside corotation of FMDWs, a wave packet 
travels away from the disk center for long-trailing and 
short-leading FMDWs; a wave packet travels toward the disk 
center for long-leading and short-trailing FMDWs. Outside 
corotation of FMDWs, the moving directions of wave packet
reverse for all these FMDW types.

The angular momentum flux carried by a FMDW is 
$$
{\cal F}_{J}=-{\pi mkr\mu_0\over (\omega-m\Omega)^2}
\bigg(C_S^2+v_c^2+C_A^2-{\pi G\mu_0\over |k|}\bigg)|
\tilde v_r|^2\ ,\eqno (4)
$$
with the surface densities of angular momentum ${\cal J}^F
\equiv m\mu_{0}|\tilde v_r|^2/[2(\omega-m\Omega)]$, of
energy ${\cal E}^F=\omega {\cal J}^F/m$ and of wave action
${\cal N}^F={\cal J}^F/m$. Inside and outside corotation,
${\cal J}^F$ is negative and positive, respectively. By 
(2) and (4), the angular momentum flux of free 
FMDWs is conserved. 
The corotation at $\omega-m\Omega(r_c)=0$ is forbidden to 
wave access by the $Q_M$-barrier when $Q_M\gsim 1$, and in 
the WKBJ regime, the {\it slightly} modified Lindblad
resonances occur at $\Gamma\equiv\kappa^2-(\omega-m\Omega)^2
+m^2C_A^2f(r)/r^2=0$ with $f(r)\sim {\cal O}(1)$ being a 
dimensionless analytic expression. Outside the 
$Q_M$-barrier, long-FMDWs exist {\it only} within the two 
modified Lindblad resonances, while short-FMDWs propagate 
within {\it and} outside the two modified Lindblad 
resonances as in the hydrodynamic case (Goldreich \& 
Tremaine 1978a).

By a massive bar, a satellite, or a large-scale spiral structure, 
an external potential $\phi^{E}(r,\theta, t)$ felt at the central 
gas disk excites FMDWs at the modified Lindblad resonances $r_M$. 
Around $r_M$, one may write $x\equiv (r-r_M)/r_M$ and 
$\Gamma\equiv {\cal G}_Mx$ with $\hbox{sgn }{\cal G}_M=\pm 1$ for 
{\it modified inner and outer Lindblad resonances} (mILR and mOLR). 
In the WKBJ regime, the set of {\it inhomogeneous} FMDW equations 
for small $x$ may be reduced to the familiar form in terms of the 
disk self-gravity potential $\phi$ as
$$
{d^2\phi\over dx^2}-i\alpha_M{d\phi\over dx}-\beta_M x\phi
=i\alpha_M\Psi\ ,\eqno(5)
$$
where 
$
\alpha_M\equiv -[2\pi G\mu_{0}r /(C_S^2+v_c^2+C_A^2)]_{r_M}
\hbox{sgn}(k),
$
$
\beta_M\equiv [r^2{\cal G}_M/(C_S^2+v_c^2+C_A^2)]_{r_M},
$
and 
$
\Psi\equiv d\phi^{E}/dx+2m\Omega\, \phi^{E}/(m\Omega-\omega).
$
The solutions of (5) involving Airy functions with proper energy
flow directions are known; the physics of these solutions studied 
previously (without magnetic field) in contexts of exciting 
large-scale galactic density waves (Goldreich \& Tremaine 1978b; 
Yuan \& Kuo 1997) and of planetary rings (Goldreich \& Tremaine 
1978a, 1982) may be extended and applied to a circumnuclear 
magnetized gas disk.
By (5), the magnitude of a FMDW is proportional to
the strength of $\Psi$ that involves $\phi^E$.

By solutions of (5), long-trailing FMDWs are dominantly 
excited by $\phi^E$ at mILR and mOLR; these waves exist
{\it between} the two 
resonances and 
propagate toward corotation after their generations at mILR 
and mOLR. Reflections at the $Q_M$-barrier transform these 
long-trailing FMDWs to short-trailing 
FMDWs that travel away from corotation and continue across the mILR 
and mOLR if dissipations were not strong. In the case of mILR, 
long-trailing FMDWs sustained by $\phi^{E}$ carries a {\it negative} 
angular momentum toward corotation; a subsequent reflection at the 
$Q_M$-barrier gives rise to short-trailing FMDWs again carrying a 
{\it negative} angular momentum that may reach the disk center 
until being disrupted by small-scale ($\lsim$ a few pc) processes 
around the nucleus.

\section{The Circumnuclear Spiral Arms and Starburst ``Ring" in NGC 1097}

NGC 1097 is a barred spiral galaxy 
with a declination of $\sim -30^{\circ}$ and 
an estimated inclination
ranging from $\sim 37^{\circ}$ to $57^{\circ}$. It is of Seyfert
type with an AGN known as a LINER 
(Barth et al. 1995) and may host a 
supermassive black hole (SMBH) of 
$\sim 2\times 10^{10}M_{\odot}$ (Rickard 1975). Estimates of 
its distance range from $11.6-17$ Mpc. 
It has a bright circumnulear starburst ``ring" 
with a diameter of $\sim 18''$ and a far infrared (FIR)
luminosity of $\sim 3\times 10^{10}L_{\odot}$ (Telesco 1988;
Gerin et al. 1988), forming stars at a 
rate of $\sim 5M_{\odot}\hbox{ yr}^{-1}$ (Hummel et al. 1987). 
The rotation curves were determined for $r\sim 4''-20''$, 
using H$_{\alpha}\ $6563\AA\hbox{ }and [N{\eightrm II}]
6584\AA \ (Meaburn et al. 1981; Blackman 1981) as well as
CO (J=$1-0$) (Gerin et al. 1988), and for $r\sim 20''-6'$,
using neutral hydrogen H{\eightrm I} 21 cm despite
a strong depletion in the bar region (Ondrechen et al. 
1989). From $r\sim 10''$ to $20''$, the inferred rotation speed
$V_{\theta}\sim 300\hbox{ km s}^{-1}$
remains roughly constant. The two dark dust lanes originating
around $\sim 4''-5''$ follow along the inner edges of the two 
spiral arms and extend almost continuously out to the bar 
(Rickard 1975) with large arm pitch angles of $\sim 50^{\circ}$ 
there. Shown in Fig. 1 is our wavelet analysis on an {\it HST} 
image of central NGC 1097 
that reveals two distinct circumnuclear spiral 
arms extended inward to $\lsim 1.5''$.

Synchrotron radio-continuum emissions (Ondrechen \& van der 
Hulst 1983;
Hummel et al. 1987) at 20 cm reveal that the intensity contour 
closely follows the dust lanes along the bar and merges 
into the central circular region around $r\sim 30''$ to $35''$. 
The total radio intensity ring overlaps with the circumnuclear 
optical/infrared ``ring" (Hummel et al. 1987; Telesco 1988), 
implying strong magnetic fields and profuse relativistic
cosmic-ray electrons. Polarized radio emission at 22, 18, 6.2 
and 3.5 cm (Beck et al. 1999) indicates that regular magnetic 
fields are parallel to the two dust lanes along the bar 
and gradually swirl into a trailing pattern in the central 
region. Relative to the circumnuclear ``ring", magnetic fields 
are inclined by $\sim 50^{\circ}$.
Despite the limited resolution, it is already indicative 
that a trailing spiral magnetic-field pattern inside the 
``ring" continues toward the nucleus. By this evidence 
and our Fig. 1, it is almost certain that spiral FMDWs (Fan \& 
Lou 1996; Lou \& Fan 1998) play a significant dynamic role in 
the circumnuclear region of a barred spiral galaxy, and magnetic 
fields provide a valuable diagnostics. The fact that in the 
``ring" of NGC 1097, the radial and azimuthal 
components of magnetic field are comparable (Beck et al. 1999) 
means that the magnetic-field swirl pattern is not so tightly 
wound there. This perhaps relates to long-trailing FMDWs that are 
excited and sustained by $\phi^E$ 
at the mILR, and spiral FMDWs involve a {\it nonaxisymmetric} 
radial magnetic field $b_r$. Fig. 1a of Beck et al. (1999) 
with 15" resolution shows a central swirl with some axisymmetry, 
while their Fig. 2 with 6" resolution shows
a central spiral pattern with nonaxisymmetry. The apparent 
central ``axisymmetric swirl" of their Fig. 1a most likely 
results from a larger beam size. 
In terms of orientations along magnetic spiral arms, 
the tendency that magnetic field of central NGC 1097 
seems to be axisymmetric instead of bisymmetric within 
the circumnuclear ``ring" region 
may be understood from the intuitive picture of distorted unaligned 
magnetic-field rings for FMDWs (Lou \& Fan 1998) similar to that of 
distorted stellar orbits for density waves (Kalnajs 1973).

For NGC 1097, the mean magnetic field $B_{\theta}$
threads through the central thin H{\sevenrm I} gas disk 
embedded with H$_2$ molecular clouds (Gerin et al. 1988). Spiral
FMDWs interact with the magnetized H{\sevenrm I} gas\footnote{A 
partial ionization and sufficiently frequent collisions between 
neutral H{\sevenrm I} gas and charged particles are presumed here.
} and H$_2$ clouds together. Random motions of H$_2$ clouds provide 
an ``effective pressure" and a necessary mechanism for FMDW damping. 
The mILR is physically identified to be located at $r\sim 10''$ 
somewhat outside the radius\footnote{For the Cassini division in 
the Saturn ring system (Goldreich \& Tremaine 1978a), particles 
outside the Mimas' 2:1 ILR have been cleared by density waves 
over 5 billion years. In the disk region outside the 
circumnuclear starburst ring of a barred spiral galaxy, there 
may be some mechanisms in operation to replenish the gas disk 
to avoid an outright gap.} 
of the circumnuclear ``ring". For $m=2$, $\Omega$ at mILR is 
$2+\sqrt{2}$ times $\Omega_p$ in a flat rotation curve, the 
corotation is located around $30''\lsim r\lsim 35''$, and 
the mOLR would be around $50''\lsim r\lsim 60''$.
The latter two estimates are only indicative, as both stellar 
and magnetized gas disks are involved at larger $r$.

Our emphasis is on FMDW processes around the mILR. Long-trailing 
FMDWs are excited and sustained at the mILR by $\phi^{E}$ and 
propagate toward corotation. They are reflected by the 
$Q_M$-barrier around corotation and travel back toward the mILR 
in the form of short-trailing FMDWs. Inside corotation, both 
long- and short-trailing FMDWs carry {\it negative} angular 
momenta (Fan \& Lou 1999). Owing to dissipation, these FMDWs 
damp and deposit in the disk their negative angular momentum
as they travel. As the disk angular momentum is persistently 
reduced, disk material outside the mILR gradually spirals inward, 
bringing along magnetic flux meanwhile. As this process persists, 
disk materials (H{\sevenrm I} and H$_{2}$ etc) and magnetic flux 
would accumulate inside the mILR. Regions of high gas density 
and enhanced magnetic flux naturally favor births of bright 
young massive stars as well as star clusters (Elmegreen 1994); 
this presumably gives rise to a circumnuclear starburst ``ring" just 
inside the mILR. Outside the mILR, the strength of a long-trailing 
FMDW should be stronger than that of a returning short-trailing FMDW 
owing to wave damping.

For order-of-magnitude estimates, we take 
$v_c\sim 35\hbox{ km s}^{-1}$ (Gerin et al. 
1988)\footnote{There seems to be no escape for a high 
Mach number of molecular clouds in the gas disk. Resonance 
processes might continuously convert FMDW energy into 
random cloud motions.} 
and $l_c\sim 100$ pc. The gross turbulent 
dissipation coefficient $\nu_c$ 
would be
$\nu_c\sim v_cl_c/3\cong 3\times 10^{26}\hbox{ cm}^2\hbox{ s}^{-1}$.
For a magnetic field of $B\sim 40\mu$G (Beck et al. 1999), a 
H{\sevenrm I} gas column density of $\lsim 10^{21}\hbox{ cm}^{-2}$ 
(Ondrechen et al. 1989), a H$_2$ gas column density 
$\sigma_{0}\sim 6\times 10^{22}\hbox{ cm}^{-2}$ (Gerin et al. 1988)
and a disk thickness $h\lsim 100$ pc, the Alfv\'en speed is
$C_A\sim 4\times 10^5\hbox{ cm s}^{-1}$. For a gas temperature
$T\lsim 100\hbox{ K}^{\circ}$, the sound speed is $C_S\lsim
10^5\hbox{ cm s}^{-1}$. 
A rule of thumb for active star 
formation in a magnetized gas disk is 
$2\pi G^{1/2}\Sigma/B\gsim 1$ with $\Sigma$ being the surface 
mass density (Shu 2001, private communications). Given the 
above parameters, $2\pi G^{1/2}\Sigma/B\sim 8$. 
The $Q_M$ estimate is uncertain, as the corotation region 
$30''\lsim r\lsim 35''$ for two-armed spiral FMDWs may involve both 
stellar and magnetized gas disks; we take $Q_M\gsim 1$. Right at the 
mILR ($r\sim 10''$), the short wavelength is 
$\lambda_S=(C_A^2+C_S^2+v_c^2)/(G\mu_{0})\sim 300$ pc, while the 
long wavelength $\lambda_L$ would be infinite; at the reflection 
point of the $Q_M$-barrier ($r\sim 30''$), they become equal 
$\lambda_S=\lambda_L=2(C_A^2+C_S^2+v_c^2)/(G\mu_{0})\sim 600$ pc.
For a mean radial wave lengthscale of $\lambda_M\sim 450$ pc, the 
damping timescale would be 
$\tau_c\sim\lambda_M^2/\nu_c\cong 6\times 10^{15}$ s. 
The timescale $\tau_w$ for a FMDW to go from the mILR to 
$Q_M$-barrier and back to the mILR is estimated to be 
$\tau_w\sim 3\times 10^{15}$ s. Thus, a sizable fraction of 
negative angular momentum carried by long-trailing FMDWs 
excited and sustained at the mILR would have been deposited 
in the gas disk as a weaker short-trailing FMDW returns to 
the mILR. As the disk angular 
momentum is reduced, gas materials with frozen-in magnetic 
flux from outside tend to accumulate just inside the mILR to 
form a circumnuclear starburst ``ring". An estimate of net mass 
inflow rate $\dot M$ involves several uncertain aspects because 
the magnitude of excited FMDWs relates to the strength of 
 $\phi^E$ and the wave damping distribution depends 
on the effective viscosity as well as some nonlinear effects. 
By taking a $|v_r|\sim 10\hbox{ km s}^{-1}$ and a
$|\omega-m\Omega|r\sim 30\hbox{ km s}^{-1}$ in 
(4), an upper limit of $\dot M$ may be estimated as
$\dot M\lsim 10M_{\odot}\hbox{yr}^{-1}$. For a stronger
$\phi^E$, nonlinear wave and damping effects may become 
important and this upper limit may be raised. 
\vskip 8.5cm

\figcaption[]{A wavelet transformed and reconstructed WFPC2/HST
(F555W filter) image in false color of the central spiral arms
around and within the circumnuclear starburst ``ring" of the
barred spiral galaxy NGC 1097 showing an angular scale lower
than 2".}

\noindent
The large pitch angle of $\sim 50^{\circ}$ of optical arms (Rickard 1975) 
and magnetic field (Beck et al. 1999) outside the ``ring" 
is interpreted in terms of long-trailing FMDWs sustained 
by $\phi^{E}$ at the mILR. As a test, it may be possible 
to detect a weaker short-trailing FMDW outside the ``ring" 
superposed with the stronger long-trailing FMDW. If one 
were to regard the large magnetic-field pitch angle as a 
non-frozen-in steady state, which we think unlikely, then 
the upper limit of $\dot M$ would be 
$\sim 1M_{\odot}\hbox{yr}^{-1}$ (Beck et al. 1999). The 
upper limit for $\dot M$ induced by turbulent viscosity
with a flat rotation curve is of the order of 
$\sim 1M_{\odot}\hbox{yr}^{-1}$. Some simulations also 
show that viscous torques are negligible with respect to 
gravitational torques several kpc away from the nucleus, 
when a relatively strong bar or tidal interaction is 
present (Combes 1991). In terms of feeding active star 
formation, the relative importance of viscous and 
gravitational torques around circumnuclear rings 
remain unresolved.

In our scenario, the effectiveness of wave damping holds 
the key for different outcomes. For a weak damping, FMDWs 
excited at the mILR eventually travel to the nucleus. For 
a strong wave damping, 
disk angular momentum is reduced primarily {\it outside} 
the mILR. For an intermediate wave damping, in addition 
to a circumnuclear ``ring" formation at the mILR, remnant 
short-trailing FMDWs also travel {\it across} the mILR
and be partially enhanced by the geometric converging 
effect (Montenegro et al. 1999). 
Damping of these remnant 
short-trailing FMDWs inside the ``ring" would induce a 
net accretion of gas and magnetic flux to fuel nucleus 
activities. For NGC 1097, in addition to the 
starburst ``ring", circumnuclear tight-winding trailing 
spiral arms of FMDWs are detected (Fig. 1) down to the 
immediate environs of the nucleus ($r\lsim 1.5''$). 
We identify this spiral structure inside the ``ring" 
(Fig. 1) as remnant short-trailing FMDWs and 
suspect that the damping of these FMDWs plays a key role in 
fueling the central SMBH on sub-kpc scales. It is highly
desirable that dedicated multiwavelength observations 
of circumnuclear regions of spiral galaxies would provide 
more examples to enrich the scenario advanced here. While 
already indicative, it is anticipated that refined maps 
of polarized radio-continuum emissions would ultimately 
establish a more specific correspondence of spiral 
structures of gas density and magnetic field around and 
within the circumnuclear starburst ``ring", analogous to 
those manifest in galactic spiral structures on larger scales. 
We expect the total and polarized radio-continuum spiral arms 
to closely follow the dust lanes along the inner edges of 
circumnuclear optical spiral arms of NGC 1097.

\acknowledgments

We thank the referee for extensive and constructive comments 
that improved this Letter. This work was supported in part 
by grants (ATM-9320357 and AST-9731623) from NSF to the 
University of Chicago and by the Visiting Scientist Programs 
at the Institute of Astronomy and Astrophysics, Academia Sinica 
(NSC-88-2816-M-001-0010-6 and NSC89-2816-M001-0006-6) and at 
the National Taiwan University (NSC89-2112-M002-037).
YQL also acknowledges the hospitality of the National 
Astronomical Observatory, Chinese Academy of Sciences.

\clearpage







\clearpage




\end{document}